# Gallium Arsenide Optical Phased Array Photonic Integrated Circuit

**MICHAEL NICKERSON[1,*], BOWEN SONG[1], JIM BROOKHYSER[2], GREGORY ERWIN[2], JAN KLEINERT[2], AND JONATHAN KLAMKIN[1]**

[1]*Department of Electrical and Computer Engineering, University of California, Santa Barbara, CA 93106, USA*
[2]*MKS Instruments, Photonics Solutions Division, 14523 SW Millikan Way, Beaverton, OR 97005, USA*
*nickersonm@ece.ucsb.edu

**Abstract:** A 16-channel optical phased array is fabricated on a gallium arsenide photonic integrated circuit platform with a low-complexity process. Tested with a 1064 nm external laser, the array demonstrates 0.92° beamwidth, 15.3° grating-lobe-free steering range, and 12 dB sidelobe level. Based on a reverse biased p-i-n structure, component phase modulators are 3 mm long with DC power consumption of less than 5 μW and greater than 770 MHz electro-optical bandwidth. Individual 4-mm-long phase modulators based on the same structure demonstrate single-sided $V_\pi{\cdot}L$ modulation efficiency ranging from 0.5 V·cm to 1.23 V·cm when tested at wavelengths from 980 nm to 1360 nm.

## 1. Introduction

Optical beam steering is increasingly investigated for emerging applications ranging from terrestrial navigation LiDAR [1] to free-space optical communication [2], climate monitoring [3], and even disaster relief [4]. Traditional beam steering methods often utilize rotating sources or gimbals with gross physical movement, resulting in physically large systems that are restricted to Hz-range steering speeds. Optical phased arrays (OPAs) resolve these mechanical limitations by providing electronic beamsteering that avoids moving parts. OPAs are simply optical versions of the familiar RF phased arrays that have been employed since the 1940s [5], forming a single large effective aperture by aligning the phase of multiple subapertures to coherently interfere their output in the far field.

Photonic integrated circuit (PIC) technology provides a scalable platform for OPAs. To date, most integrated OPA research has focused on silicon photonics (SiPh) and operation near 1550 nm [6–11]. Shortcomings of SiPh technology include the use of phase modulators based on thermally-induced effects that are limited to MHz modulation speeds with high power dissipation, or carrier-based phase modulators which are faster but suffer from high residual amplitude modulation (RAM) [12–15]. The silicon bandgap also limits operation to wavelengths greater than 1100 nm, thereby excluding the region near 1000 nm commonly utilized for topographical and remote-sensing LiDAR [16–18]. Additionally, on-chip gain and lasers for SiPh are only possible with complex integration techniques involving flip-chip or wafer bonding. In contrast, group III-V compound semiconductor platforms offer native optical gain, lasers, and efficient low-RAM phase modulators.

OPAs based on III-V compound semiconductors have primarily focused on indium phosphide (InP) PIC platforms in the 1300 nm and 1550 nm telecommunications wavelength regions. Low power OPAs have been demonstrated on these platforms, some with on-chip gain [19–25]. However, InP PICs are not suitable for shorter wavelengths and typically have high manufacturing costs.

Gallium arsenide (GaAs), another III-V material, has long been employed to build discrete optical components [26–29]. While the development of PIC technology has been focused on the InP system due to its support for common telecommunications wavelengths, GaAs fabrication methods have also been matured by the wide use of this material in high power microwave electronics and diode lasers operating near 900 nm.

In this work, we developed a low fabrication complexity GaAs PIC platform and leveraged it to demonstrate an OPA with 16 channels, illustrated in Figure 1. The integrated phase modulators exhibited broad bandwidth, high speed, and low RAM. This demonstration does not include on-chip optical gain, but the platform is capable of incorporating gain in the 900-1300 nm range without sacrificing OPA performance [30–32].

With a single input and an array of phase modulators collapsing to a dense 4-µm-pitch edge-coupled output, this 16-channel PIC OPA achieves a 0.92° beamwidth with 15.3° grating-lobe-free steering range and 12 dB sidelobe level when characterized with a 1064 nm external laser. When tested with a 1030 nm external laser, individual 3 mm long phase modulators demonstrate single-sided $V_\pi \cdot L$ efficiency of 0.7 V·cm and RAM below 0.5 dB for greater than $4\pi$ phase modulation [33], DC power consumption of less than 5 µW at $2\pi$ modulation depth, and greater than 770 MHz electro-optical bandwidth when mounted on and wire bonded to a printed circuit board (PCB).

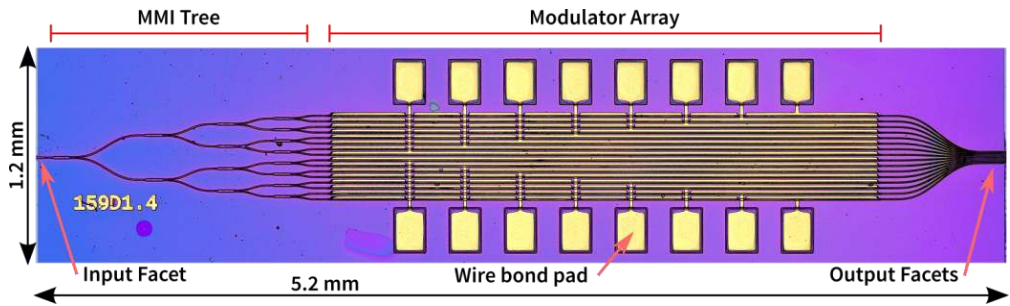

**Fig. 1.** Optical micrograph of fabricated PIC.

## 2. PIC platform design

### 2.1 PIC architecture

Figure 1 shows a fabricated PIC with a footprint of 5.2 mm × 1.2 mm. The input is a single 5 µm wide ridge waveguide with a cleaved facet, designed for high coupling efficiency from a lensed fiber. The waveguide then tapers to a width of 2 µm and is split into 16 identical channels via a cascaded tree of 1×2 multi-mode interference (MMI) splitters. Each channel comprises identical 3-mm-long, 2-µm-wide, low-power, low-RAM electro-optic phase modulators. The array then collapses to a dense lateral pitch of 4 µm at the output cleaved facet. The PIC was limited to 16 channels to simplify integration and testing, but is readily scalable to significantly higher channel count.

This 16-channel PIC was specifically designed as an OPA demonstrator for this GaAs platform, suitable for a broad range of center wavelengths and continuously usable throughout typical tunable laser ranges. No gain material is present, but the components and epitaxy are engineered for compatibility and simple monolithic integration of conventional GaAs-based quantum wells or quantum dots in the 880-1300 nm range [31,32,34]. To tailor the design for specific wavelengths of interest, appropriate gain structures would be developed, and slight adjustments to MMI splitter geometry would be made if operating far from the 1030 nm wavelength utilized in this work.

### 2.2 Phase modulators

Careful attention was paid to develop phase modulators ideally suited for OPA applications. Low propagation loss is critical for channel count and output power scaling, and low RAM is necessary for continuous beamforming over the entire steering range. Low power consumption is important for compact portable applications, and high electro-optical bandwidth is required for rapid beam scanning.

To satisfy these requirements, the modulators were developed as a GaAs-based reverse-biased P-p-i-n-N double heterostructure, detailed in Table 1 along with the refractive indices and assumed material losses used for simulation. This structure yields a high overlap between a nearly linear electric field along the [001] direction and the optical mode as shown in Figure 2(a), efficiently utilizing the strong linear and quadratic electro-optic effects of GaAs for TE modes propagating in the [01$\bar{1}$] direction [35–37]. Additional modulation is possible with free carrier effects, but these effects are avoided to reduce RAM. Figure 2(b) reports the expected phase modulation contributions at 1030 nm for a modulator based on this design. Low optical absorption and RAM are achieved by realizing an 85% overlap between the optical field and the GaAs guiding region, and only 19% overlap with free carriers at a -10 V bias.

**Table 1. Phase modulator epitaxial layer structure**

| Material | Thickness [nm] | Doping [cm$^{-3}$] | Refractive Index | Material Loss [dB/cm] |
|---|---|---|---|---|
| GaAs | 300 | (p) 1e19 | 3.4653 | 140 |
| Al$_{0.2}$Ga$_{0.8}$As | 1000 | (p) 1e18 | 3.4046 | 15 |
| Al$_{0.4}$Ga$_{0.6}$As | 200 | (p) 4e17 | 3.3551 | 19 |
| Al$_{0.4}$Ga$_{0.6}$As | 200 | (p) 2e17 | 3.3551 | 11 |
| GaAs | 100 | (p) 2e17 | 3.4653 | 7.2 |
| GaAs | 500 | UID | 3.4653 | 0.04 |
| GaAs | 100 | (n) 2e17 | 3.4653 | 8.0 |
| Al$_{0.3}$Ga$_{0.7}$As | 200 | (n) 2e17 | 3.3811 | 13 |
| Al$_{0.2}$Ga$_{0.8}$As | 400 | (n) 4e17 | 3.4046 | 21 |
| Al$_{0.2}$Ga$_{0.8}$As | 1000 | (n) 2e18 | 3.4046 | 100 |
| GaAs | 300 | (n) 3e18 | 3.4653 | 120 |
| GaAs | substrate | (n) 1e18 | 3.4653 | 40 |

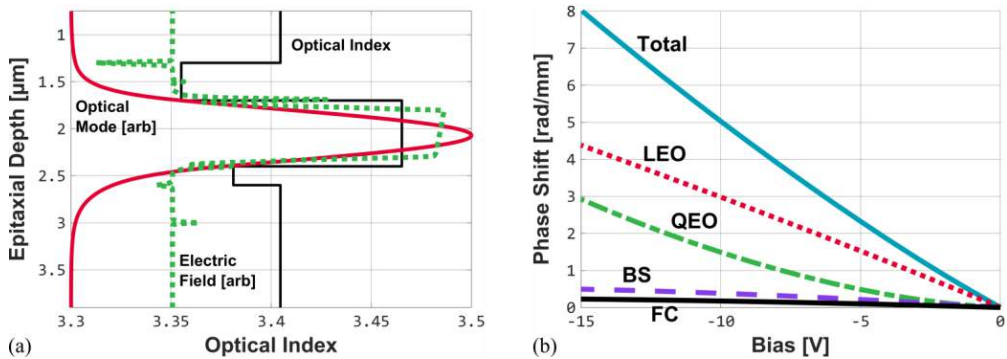

**Fig. 2.** 1D simulations showing (a) optical index (black), optical mode (red) and electric field (green, dotted), (b) phase modulation at 1030 nm (cyan) including contributions from linear electro-optic (LEO, red, dotted), quadratic electro-optic (QEO, green, dot-dashed), bandgap shift (BS, purple, dashed), and free carrier plasma (FC, black) effects.

The phase modulators were physically implemented as 2-µm-wide deep ridge waveguides with P-contacts deposited on top and an N-contact applied to the backside of the thinned, conductive substrate. A 2.5-µm-thick passivation and isolation material composed of alternating layers of silicon dioxide (SiO$_2$) and silicon nitride (SiN) is present over the entire PIC. Vias are formed on top of the phase modulator ridges to access the P-contacts. Cross-modulator routing is achieved by metal deposited on top of the isolation material, and along-modulator conductivity is assured with a buried metal layer present only over the modulators.

The left half of Figure 3 shows a diagram of the modulator cross sections and cross-modulator routing.

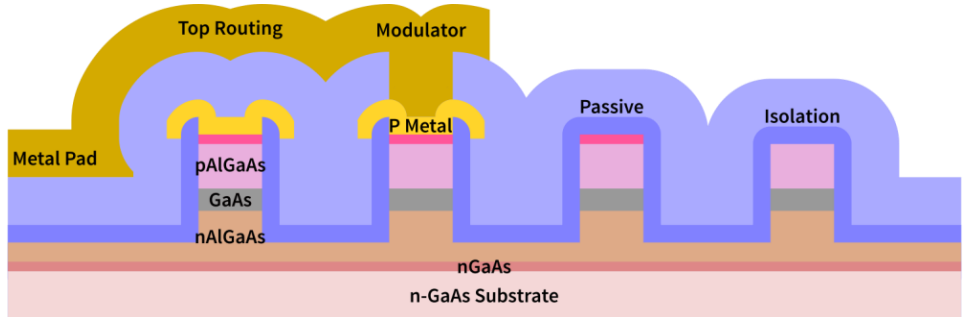

**Fig. 3.** Cross-section schematic of epitaxy and assorted waveguide structures.

### 2.3 Passive components

The remainder of the PIC consists of assorted passive components formed by the single deep ridge etch that is shared with the phase modulators. Passive 2-µm waveguides route the light and have the same structure as modulators, omitting the metal contacts and top vias. Cleaved waveguide facets with antireflection (AR) coating form the single input and 16 outputs. To effect longitudinal electrical isolation between modulators and other passive components, short segments of waveguides on either end of modulators are turned into electrical isolation regions by removing the top p-GaAs layer. The right half of Figure 3 illustrates these passive components, and Figure 4(a) shows the simulated fundamental TE mode.

A 2 µm waveguide width was chosen for fabrication simplicity and lithographic resolution constraints, but is not inherently single mode. Single-mode operation is weakly achieved from the stronger bending loss and MMI splitter excess loss experienced by higher order modes. Strict single-mode operation requires waveguides roughly 0.5 µm wide, which may be fabricated with higher resolution lithography.

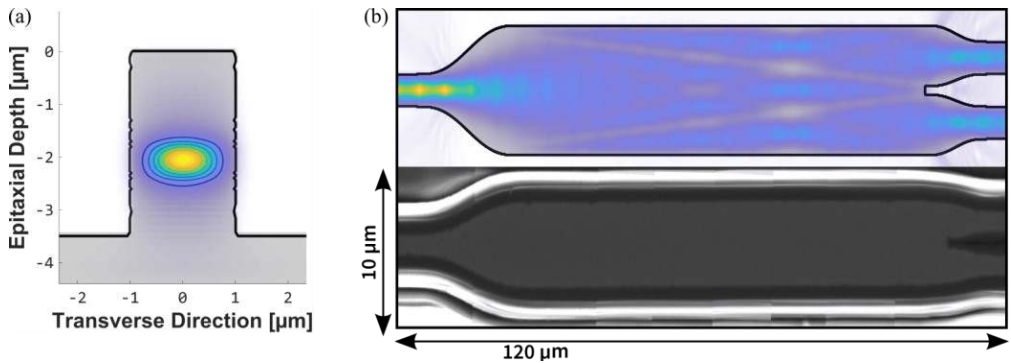

**Fig. 4.** (a) Simulated fundamental TE optical mode of modulator and passive waveguides (color) and waveguide structure (shaded grey). (b) Simulated (top) and SEM (bottom, light grey, surrounded by bright and dark dielectric material) image of 1×2 MMI splitter.

The MMI splitters forming the 1×16 splitter share the same passive waveguide structure, with geometry initially determined by standard high contrast MMI ratios [38]. 3D FDTD simulation of the epitaxial structure was performed with commercial tools [39] and used to slightly modify the geometry for optimal performance, as shown in Figure 4(b).

Center-fed 1×2 MMI splitters were selected as an element of the broadband design, and have simulated excess optical loss under 3 dB from 920 nm to 1280 nm. 1×4 MMI splitters were also successfully implemented with a 280 µm shorter tree, but are not as broadband (1000-1180 nm); the slight length savings may be traded off against further geometry optimization for specific wavelengths.

## 3. PIC fabrication

The fabrication process was designed for low fabrication complexity, and consists of seven mask sets, two GaAs etches, and two metal lift-off steps. Figure 5 shows schematic diagrams of the primary fabrication process steps.

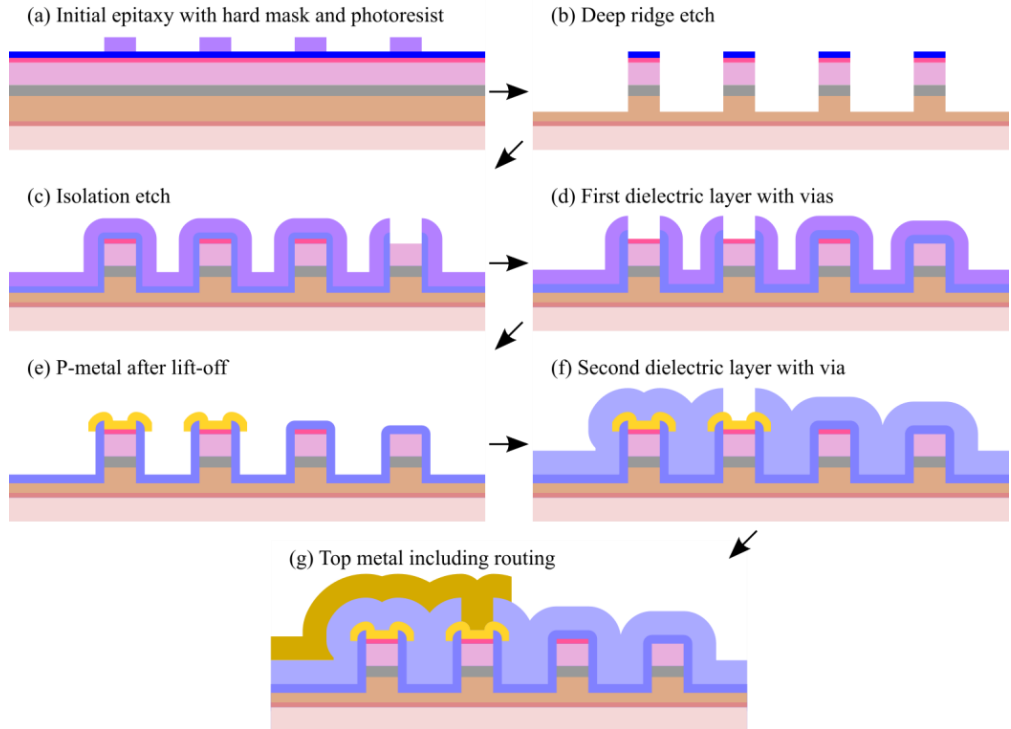

**Fig. 5.** Schematic diagrams of primary fabrication process steps. (a) Initial epitaxy with hard mask (blue) and patterned photoresist (purple). (b) Deep ridge etch. (c) Isolation etch. (d) First dielectric layer (light blue) with p-via openings. (e) P-metal (yellow) after lift-off. (f) Second dielectric layer (lighter blue) with via opening. (g) Top metal (dark yellow) including routing.

The process begins with the deposition of a 400 nm $SiO_2$ hard mask on clean epitaxy via plasma enhanced chemical vapor deposition (PECVD). Photoresist (PR) is patterned on top of this by an i-line stepper with 0.8 µm effective resolution, defining the waveguide pattern. Unprotected regions of the hard mask are removed by inductively coupled plasma reactive ion etching (ICP-RIE) with $CHF_3/CF_4$ chemistry, and the waveguides are defined by a 4 µm deep $Cl_2/N_2$ ICP-RIE etch. Figure 6(a) shows the sidewall of a waveguide immediately after etching. After removing the waveguide hard mask with buffered HF, another hard mask is deposited and patterned, and a $BCl_3/SF_6/N_2$ selective ICP-RIE etch is used to remove the top p+GaAs layer in the electrical isolation regions without etching the top AlGaAs cladding. The isolation-etch hard mask is removed and the devices are passivated with 5 nm $Al_2O_3$ and 10 nm $SiO_2$ by atomic layer deposition, on top of which the first dielectric isolation layer of 550 nm total thickness is deposited in alternating $SiO_2$ and SiN layers via PECVD.

The first dielectric isolation layer is selectively opened above the modulator waveguides by a PR mask and $CHF_3/CF_4$ ICP-RIE etch, and a lift-off mask is defined with negative PR. After depositing 20/40/500 nm of Ti/Pt/Au with electron beam deposition, the metal is lifted off to define the lower metal layer, forming the modulators' top metal contacts. Another PECVD step forms the second dielectric isolation layer with 2 μm total thickness of alternating SiO$_2$ and SiN layers, and vias are opened to the first metal layer by another ICP-RIE etch on a PR mask. The second lift-off step with 3 μm of Ti/Au then forms the top metal layer for routing and wire bond pads. Shown in Figure 6(b) is a cross-modulator trace formed by this second lift-off step. A final lithography and ICP-RIE etch defines cleave lanes by removing all dielectric layers from selected areas. To complete fabrication, the substrate is thinned to approximately 200 μm, 10/50/100/20/500 nm of Ni/Ge/Au/Ni/Au N-metal is deposited on the back side, and devices are rapidly thermally annealed in forming gas at 400°C then 480°C for 10 seconds each.

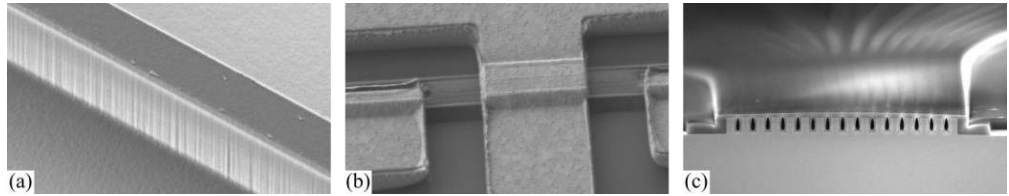

**Fig. 6.** SEM images of selected fabrication process steps. (a) Deep ridge after first etch. (b) Top metal cross-modulator routing. (c) Output facet after cleaving and AR coating.

After electrical testing to determine initial yield, selected dies are separated, facets are formed by mechanical cleaving, AR coatings are applied to the facets, and individual PICs are singulated. Figure 6(c) shows a typical cleaved and AR coated output facet, and Figure 7(a) illustrates the scale of a singulated 16-channel PIC. Individual PICs are then mounted to aluminum nitride carriers with electrical traces and wire bonded. The PIC-on-carriers are then soldered and wire bonded to PCBs as shown in Figure 7(b).

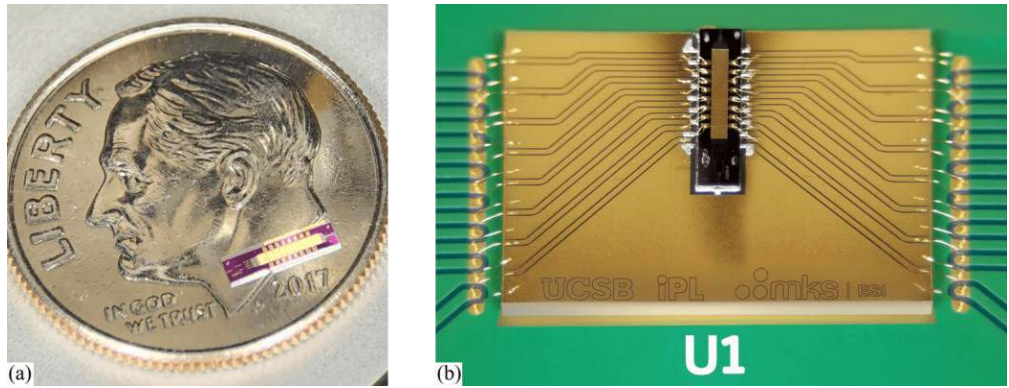

**Fig. 7.** (a) Photograph of 16-channel PIC on US dime. (b) Photograph of PIC-on-carrier on PCB, retouched for clarity.

## 4. Characterization

### 4.1 Phase modulator optical performance

Optical measurement is performed by edge coupling a lensed fiber into the cleaved-facet PIC input, with the PIC mounted on a thermoelectric cooler kept at 20°C. Packaged 980 nm and 1030 nm fiber-coupled lasers and an O-band tunable laser provide the light.

Phase modulators are characterized primarily with single-sided Mach-Zehnder modulator (MZM) test PICs comprising a single input, a 1×2 MMI, a 4 mm phase modulator on one arm, a 2×1 MMI, and a single output. The electro-optic modulation response is found by rapidly sweeping the bias and recording the simultaneous bias and MZM transmission waveforms on a digital signal oscilloscope, then transforming these to transmission as a function of bias by matching timeseries data. An accurate estimate of the phase modulation ($\Delta\phi$) and RAM ($\Delta\alpha$) as a function of bias is obtained by simultaneously fitting multiple normalized measurements to the analytical power transmission ($T$) of a single-sided unbalanced-loss MZM:

$$E_1 = (1-k)\, E_0;\; E_2 = k\, E_0\, e^{i\,\Delta\phi - \Delta\alpha/2} \tag{1}$$

$$\begin{aligned} T/E_0^2 &= |E_1 + E_2|^2\, F + (1-F) \\ &= 1 + F\, k\, (-2 + k + k\, e^{-\Delta\alpha}) - 2\, F\, (k-1)\, e^{-\Delta\alpha/2} \cos(\Delta\phi) \end{aligned} \tag{2}$$

where $k$ is the splitting ratio and $F$ is the coherent fraction. To reduce the degrees of freedom, the phase modulation and RAM are empirically parameterized as functions of bias:

$$\Delta\phi(V) = p_1\, (V_0 - V) + p_2 (V_0 - V)^2 \tag{3}$$

$$\Delta\alpha(V) = a_1\, (\exp[a_2(V_1 - V + dV)])\, (1 - \tanh[V - V_1 + dV])/2 \tag{4}$$

where $V_0 = 1.62$ V represents the intrinsic electrical field across the p-i-n junction with a value chosen from electrical simulations, and $dV = 1$ V is a manually chosen smoothing offset. Parameters $V_1$, $p_1$, $p_2$, $a_1$, and $a_2$ are fit at each measured wavelength.

Tested at wavelengths from 980 nm to 1360 nm, the 4 mm phase modulators in the MZM test PICs demonstrate single-sided V$_\pi$·L modulation efficiency ranging from 0.5 V·cm to 1.23 V·cm for the TE optical mode, as shown in Figure 8(a). RAM is under 0.5 dB until the onset of Franz-Keldysh electroabsorption, reaching 3 dB at a reverse bias of 5.9 V at 980 nm (>5π rad), 12 V at 1030 nm (>8π rad), and equipment-limited >15 V at 1260 nm (>6π rad) and longer.

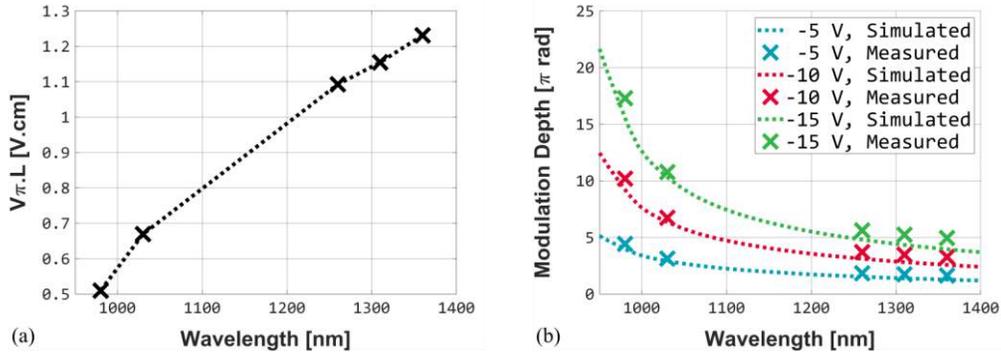

**Fig. 8.** (a) 4 mm phase modulator efficiency at various wavelengths. (b) Simulated (lines) and measured (crosses) phase modulation of 4 mm device at various wavelengths and biases.

These MZM results are verified at 1030 nm by measurements of individual 3 mm modulators on the 16-channel PICs, performed with a self-heterodyne technique in which the modulator is treated as a single arm of a fiber-optic MZM while the other arm is offset by 150 MHz with an acousto-optic modulator (AOM) [40]. Figure 9 shows a diagram of this arrangement. The laser output is split with a fiber-optic splitter, one output is coupled into the PIC and out of one modulator, and the other output is frequency shifted with a fiber-coupled AOM. After recombining in a fiber-optic splitter, the output of the photodiode is equivalent to a phase-shift-keyed (PSK) signal with a carrier frequency equal to the AOM shift and the phase shift equal to the phase applied by the modulator. The PSK waveform can then be decoded with standard IQ demodulation techniques to extract the relative phase shift. By dithering the applied modulation bias with a small amplitude around a DC offset, precise measurements of the phase modulation's first and second derivative and the first derivative of RAM with respect to bias can be obtained at any AC modulation frequency that is less than half the AOM frequency.

By simulating the electric and optical fields of these modulators with commercial solvers [39] and applying well-known literature values for the various GaAs phase modulation effects [35–37,41,42], the expected performance of these modulators is calculated. Without adjusting any model values, the expected performance shown in Figure 8(b) is in very good agreement with measurements across the full bias and wavelength range. The slight underestimate may be due to free carrier effects, especially at longer wavelengths where the electro-optic effects are weak, as literature coefficients for free carrier contributions are inconsistent.

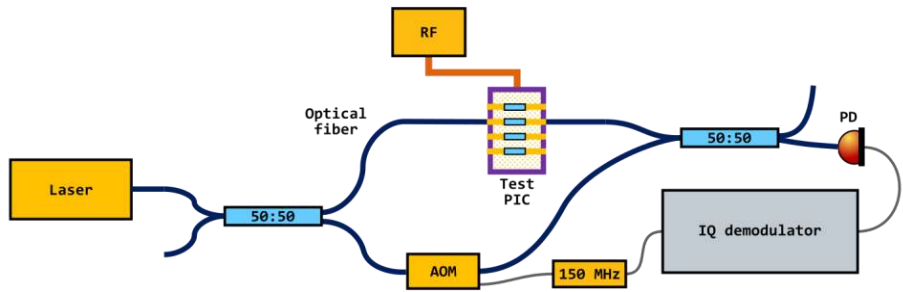

**Fig. 9.** Diagram of self-heterodyne measurement technique.

### 4.2 Phase modulator electrical performance

Detailed input-reflection (S11) electrical measurements are performed with a network analyzer on several PCB-mounted PICs. The expected electrical transmission (S12) and relative contributions of the packaging and PIC are determined by separately characterizing the packaging PCB and applying a lumped-element model to fit the resistance, capacitance, and inductance of the PCB, carrier, PIC, and wire bonds to the S11 measurements. From this, the PIC modulators are determined to have $5 \pm 3\ \Omega$ series resistance and $3.6 \pm 0.2$ pF capacitance. The expected 3 dB electrical bandwidth of the phase modulators on mounted PICs is 700 MHz, and minor changes to the PCB are identified that should increase this to 1.4 GHz.

Direct electro-optical responsivity is characterized by modulating a single channel and measuring the photodiode response at a selected point in the far-field output, demonstrating a 6 dB bandwidth between 770 MHz and 1 GHz for several channels of a selected PIC, as shown in Figure 10(a). This agrees with the simulated electrical bandwidth, as power modulation relates to the square of the phase modulation for small signals.

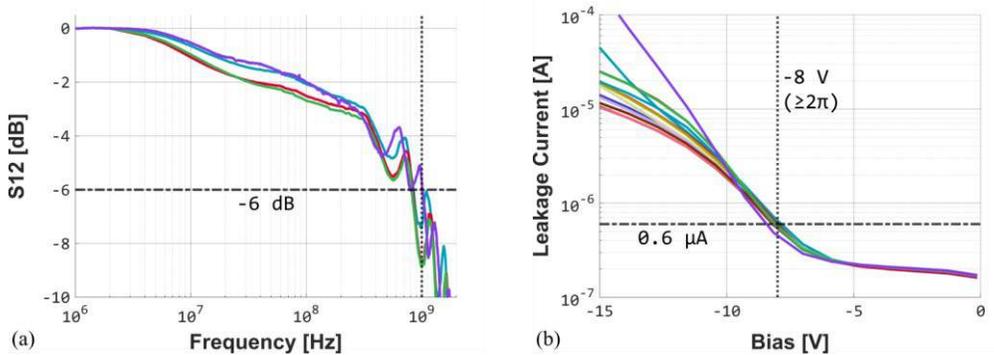

**Fig. 10.** (a) Electro-optical response of 4 channels of a selected PIC. (b) DC leakage current under reverse bias for all 16 channels of a selected PIC.

The modulation bandwidth of these 3 mm long phase modulators is ultimately limited by their capacitance, and is thus reducible by shortening the modulators. At 1030 nm, modulators only 1.2 mm long could provide a full $2\pi$ phase shift with 1 dB RAM. GaAs is also amenable to traveling-wave electrode designs exceeding 40 GHz bandwidth [43–45].

DC electrical testing of the 3 mm modulators, plotted in Figure 10(b), shows leakage currents under 0.6 µA at -8 V bias, which corresponds to more than $2\pi$ modulation depth at all tested wavelengths and is consistent across all modulators of a selected PIC. Operating up to 1360 nm, the full 16-channel PIC thus requires under 100 µW to hold any steering angle.

As indicated in Table 2, these optical and electrical metrics are competitive with most InP, SiPh, and lithium niobate on insulator (LNOI) modulators from literature, demonstrating higher speed and lower power than SiPh thermal modulators, lower RAM and power than carrier-injection modulators, and higher efficiency than LNOI modulators.

**Table 2: Phase modulator performance comparison**

| Year | Platform | Wavelength [nm] | Method[a] | Length [mm] | $V_\pi \cdot L$[b] [V·cm] | RAM @ $2\pi$ [dB] | Power @ $2\pi$ | EO Bandwidth |
|---|---|---|---|---|---|---|---|---|
| This work | GaAs | 980-1360 | EO | 4 | 0.5-1.23 | <0.5 | <5 µW | 1 GHz |
| 2020 [46] | GaAs | 980 | EO | 5 | 1.15 | - | - | 20 MHz |
| 2014 [47] | GaAs | 780 | EO | 2 | 1.5 | - | - | 13 MHz |
| 2013 [48] | GaAs | 1550 | EO | 7 | 0.42 | 1 | - | - |
| 2008 [49] | GaAs | 1550 | EO | 7 | 0.3 | <1 | - | - |
| 1998 [50] | GaAs | 1310 | EO | 2 | 0.52 | ~3 | - | 2.5 GHz[c] |
| 1991 [67] | GaAs | 1064 | EO | 1.8 | 1.9 | - | - | 2 GHz[c] |
| 1990 [41] | GaAs | 1090 | EO | 5.3 | 1.3 | ~3 | - | - |
| 2021 [51] | InP | 1550 | carrier | 0.8 | 0.2 mA·cm | - | 48 mW | 1 GHz[c] |
| 2013 [24] | InP | 1520-1570 | carrier | 0.2 | 0.2 mA·cm | 1.5 | 20 mA | 100 MHz |
| 2023 [19] | InP | 1550 | EO | 3.5 | 1.02 | - | 200 µW | - |
| 2022 [21] | InP | 1480-1550 | EO | 2.5 | 1.2 | <0.5 | - | - |
| 2014 [44] | InP | 1550 | EO | 10 | 1.6 | - | - | 67 GHz |
| 2020 [20] | InP | 4600 | thermal | 2.35 | 53 mW·cm | - | 450 mW | - |
| 2023 [52] | LNOI | 850 | EO | 5.5 | 1.56 | - | - | >20 GHz |
| 2023 [52] | LNOI | 1550 | EO | 5.5 | 2.58 | - | - | >40 GHz |
| 2023 [53] | LNOI | 400-700 | EO | 8 | 0.34-0.96 | - | - | >20 GHz |
| 2021 [54] | LNOI | 1064 | EO | 7 | 3.82 | - | - | - |
| 2018 [55] | LNOI | 1550 | EO | 5 | 4.4 | 0.1 | - | 100 GHz |
| 2019 [56] | LNOI + SiPh | 1550 | EO | 3 | 4.4 | - | - | 70 GHz |
| 2018 [57] | LNOI + SiPh | 1550 | EO | 5 | 13.4 | - | - | 106 GHz |
| 2020 [14] | SiPh | 1550 | carrier | 0.5 | 0.03 | 5 | 34 mW | 324 MHz |
| 2013 [58] | SiPh | 1550 | EO | 0.75 | 2 | 3.2 | - | 27 GHz |
| 2008 [59] | SiPh | 1550 | EO | 1 | 6 | 2.8 | - | 30 GHz |
| 2022 [6] | SiPh | 1550 | thermal | 1.5 | 0.7 mW·cm | - | 9 mW | 18 kHz |
| 2020 [60] | SiPh | 488 | thermal | - | - | - | 30 mW | 50 kHz |
| 2020 [61] | SiPh | 1560 | thermal | 0.5 | 0.8 mW·cm | - | 31.5 mW | 200 kHz |
| 2019 [12] | SiPh | 1550 | thermal | - | - | 2.4 | 5 mW | 10 kHz |
| 2018 [62] | SiPh | 1550 | thermal | 0.44 | 1.1 mW·cm | - | 48 mW | - |
| 2014 [9] | SiPh | 1480-1580 | thermal | - | - | ~0 | 20 mW | 7.3 kHz |
| 2011 [63] | SiPh | 1555 | thermal | - | - | - | 430 mW | - |

[a] "EO" = electro-optic reverse biased, "carrier" = carrier injection, "thermal" = thermo-optic effect
[b] Single-sided; doubled where push-pull is reported. [c] Not measured, expected RC limit reported.

*4.3 Optical phased array performance*

Key metrics for OPAs are the full-width half-maximum beamwidth – the grating-lobe-free steering range within which the beam can be unambiguously directed without confusion with a grating lobe – and the sidelobe level (SLL) or background level. From the beamwidth and steering range, the number of unique addressable points can be computed, which is a scale-free metric that does not depend on the linear transformations possible with optical lens systems. The number of addressable points is approximately the number of channels for dense linear arrays with ideal beam quality; sparse arrays with nonuniform offsets between channels are able to trade worsened SLL for more addressable points.

Analytically, the steering range can be estimated by $\arcsin(\lambda/d)$ and the beamwidth by that of an aperture equivalent to the OPA width, $\sim\lambda/D$, where $\lambda$ is the free-space wavelength, $d$ is the distance between emitters, and $D$ is the width of the OPA. The envelope of the steered lobe is determined by the single-emitter beam pattern.

In the case of this PIC tested at $\lambda = 1064$ nm, $d = 4$ μm, and $D = (16-1) \cdot d = 60$ μm, the expected steering range is 15° and expected beamwidth is ~1°, and the number of addressable points is ~15. These results are confirmed by directly propagating the simulated output facet optical mode to the far field [64]; the simulated steering range is 15.2° with a 0.85° beamwidth for ~17 addressable points. The SLL of 13 dB is limited by the $sinc^2$ function of the far-field pattern of a uniform aperture [65], and cannot be reduced without nonuniform emitter spacings or output powers.

The phased-array performance and output pattern of the selected PIC is evaluated using the experimental setup shown schematically in Figure 11. The PIC-on-carrier mounted on a PCB is secured to an optical bench, and a lensed fiber on a 3-axis fiber alignment stage couples a 1064 nm laser into the PIC's input facet. An in-line fiber polarizer adjusts the input as needed for optimal polarization. The output of the PIC is captured by a 4 mm effective focal length (EFL) objective lens with a 0.65 numerical aperture, and then a 500 mm EFL relay lens re-images the emitter plane with a 500 μm effective emitter spacing.

Past the reimaged emitter plane, the beam is split with a 3° wedge prism; one surface reflection travels through a pair of relay lenses that reproduce the intermediate focal plane on a monitoring camera. The other surface reflection is propagated to the far field and collimated onto an Ophir LT665 beam profiling camera with a 1147 mm EFL lens, while an identical lens images the transmitted far field beam onto a photodiode. The beam profiling camera may be moved to the near field position for near field data collection.

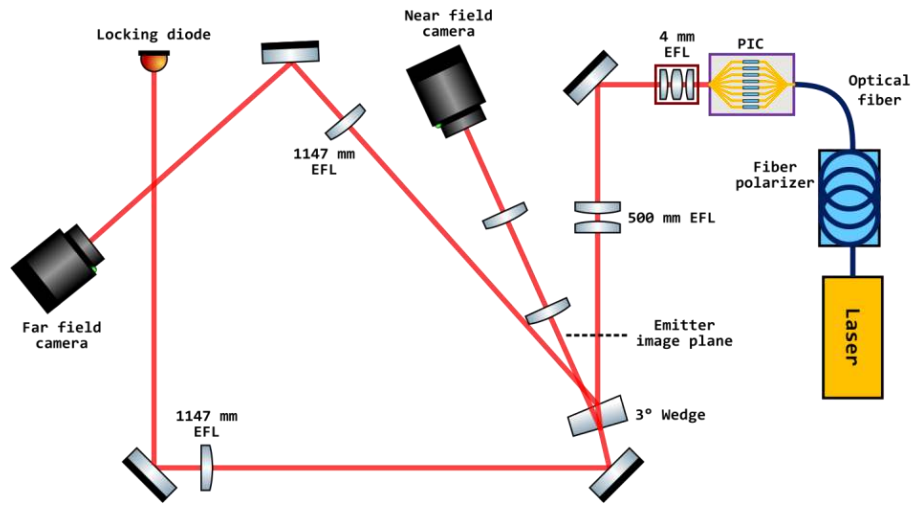

**Fig. 11.** Schematic of OPA optical measurement setup. Not to scale.

The output phase of each channel is offset by an unknown amount due to differing path lengths and fabrication imperfections, so the LOCSET algorithm [66] is used to cohere the OPA output by applying the requisite phase modulation to each channel to maximize the signal on the far-field photodiode. To steer the output, the photodiode is moved on a micrometer, and the LOCSET algorithm adjusts emitter phases to shift the peak of the beam to the new location. The positional beam profile is converted to an angular profile at the OPA output by backing out the optical transformations: $\theta_{OPA} = \frac{x_{camera}}{1147 \text{ mm}} \cdot \frac{500}{4}$. Data is captured at the 4.54-μm pixel pitch of the beam profilometer, corresponding to $0.028°$ angular resolution at the OPA output.

Figure 12(a) shows the measured beam profile at $0°$ steering compared to expected performance. Measured beamwidth is $0.92°$ with $15.3°$ steering range and 12 dB SLL, differing from simulation by only 8%, 0.7%, and 1 dB respectively. Capturing further beam profiles at additional steering angles, shown in Figure 12(b), the PIC output reproduces the expected envelope and maintains performance with $0.93° \pm 0.01°$ beamwidth and $12.1 \pm 0.4$ dB SLL across the entire grating-lobe-free steering range as indicated in Figure 12(c).

The slight difference between expected and measured steering range may be due to minor imprecision in the measurement optics, pattern scale errors during fabrication, or a combination of both. Fabrication scale error is expected to be under 1%. The higher beamwidth and inferior SLL can be partially attributed to slightly mismatched subaperture phases, as indicated by the lack of clearly resolvable uniform sidelobes, and nonuniform channel powers.

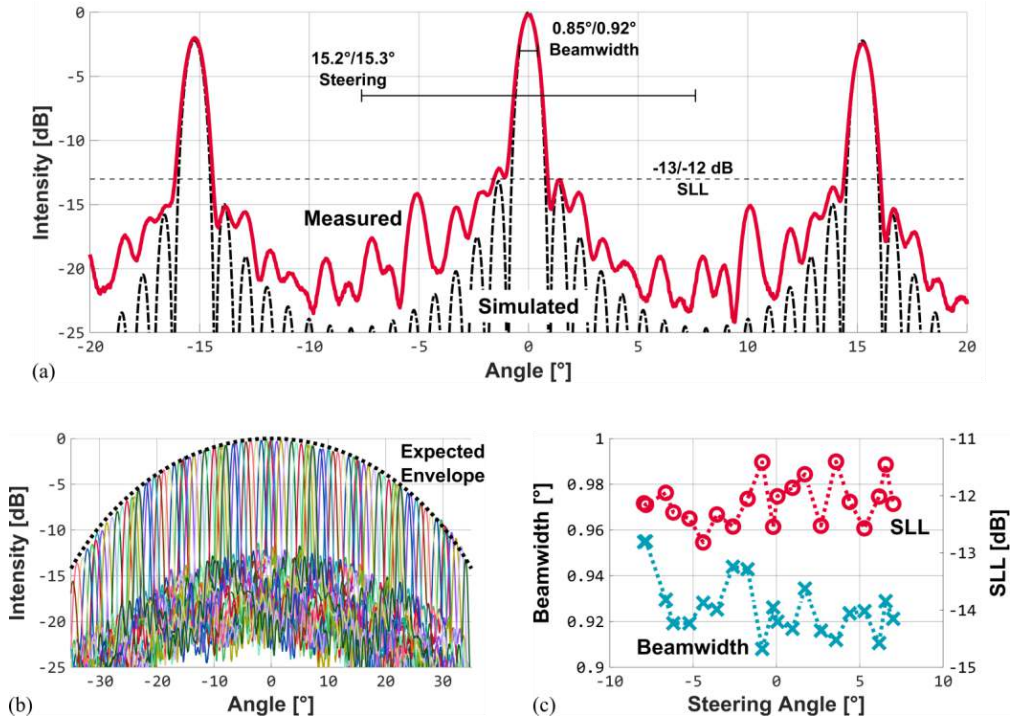

**Fig. 12.** (a) Simulated (black, dashed) and measured (red, solid) beam profiles with indicated simulated/measured beam width, steering range, and sidelobe level. (b) Measured beam profiles at 21 different steering angles (colored, solid), compared to expected envelope (black, dotted). (c) Beamwidth (left, crosses) and sidelobe level (right, circles) at measured steering angles.

The near field beam profile at zero bias is also captured. A typical output image, Figure 13(a), shows that power is not evenly distributed amongst all channels. The uniformity exhibits dependence on input fiber position; comparing the per-channel output power across 12 different input alignments in Figure 13(b), the mean emitter-to-emitter standard deviation across the array is 1.2 dB, and the mean deviation of a single emitter at different input alignments is 0.76 dB. The variation with input alignment and excess power in the center, taken together, indicates likely multi-mode coupling into the MMI tree.

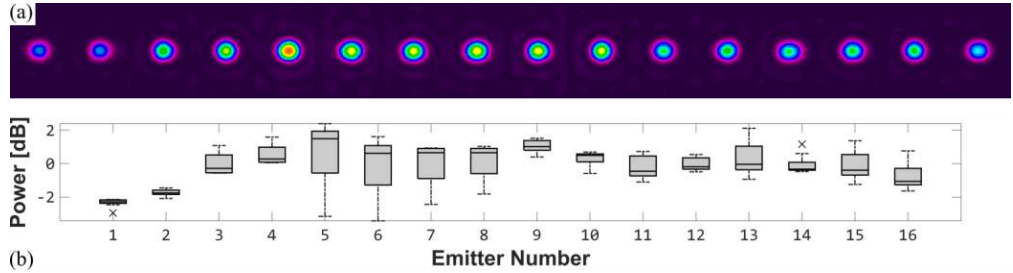

**Fig. 13.** (a) Typical image of PIC output facets. (b) Distribution of per-channel output power for 12 separate input fiber alignments with median (line), 25th to 75th percentile limits (box), 10th to 90th percentile limits (whiskers), and outliers (crosses).

The performance of this OPA is comparable to or better than most other results in the literature, as summarized in Table 3. The SLL of 12 dB is close to the theoretical 13 dB ultimate performance of a uniform linear array and better than the majority of reported results. Beam quality for a uniform array can be judged by comparing the number of addressable points to the number of channels, and this PIC compares favorably to other uniform arrays. Nonuniform arrays can achieve higher addressable points by sacrificing SLL. This PIC demonstrates operation at 1064 nm, which has not been widely demonstrated for OPAs, and is attractive for expanding the potential applications of OPAs.

**Table 3: Optical phased array performance comparison**

| Year | Platform | Wavelength [nm] | Channels | Steering Range [°] | Beamwidth [°] | Addressable Points | SLL [dB] |
|---|---|---|---|---|---|---|---|
| This work | GaAs | 980 – 1360[a] | 16 | 15.3 | 0.92 | 16 | 12 |
| 2020 [46] | GaAs | 980 | 15 | 30 | 4.7 | 6 | 8 |
| 1991 [67] | GaAs | 1064 | 10 | 20 | 2 | 10 | - |
| 2023 [19] | InP | 1550 | 32 | 35 | 0.46 | 76 | 8.2 |
| 2022 [23] | InP | 1550 | 30 | 17 | 1.49 | 11 | - |
| 2022 [21] | InP | 1480 – 1550 | 8 | 17.8 | 2.5 | 7 | 12.8 |
| 2021 [51] | InP | 1550 | 100 | 8.88 | 0.11 | 80 | 6.3 |
| 2020 [20] | InP | 4600 | 32 | 23 | 0.6 | 38 | - |
| 2013 [24] | InP | 1520-1570 | 8 | 1.7 | 0.2 | 8 | 10 |
| 2023 [68] | SiPh | 1550 | 16 | 120 | ≥6.6[b] | <18[b] | 10.9 |
| 2022 [8] | SiPh | 1550 | 8192 | 100 | 0.01 | 10000 | 10 |
| 2022 [6] | SiPh | 1550 | 256 | 45.6 | 0.154 | 296 | 10.8 |
| 2020 [61] | SiPh | 1560 | 32 | 18 | 0.63 | 28 | 10 |
| 2020 [60] | SiPh | 488 | 64 | 50 | 0.17 | 294 | 6.05 |
| 2020 [69] | SiPh | 1525-1600 | 512 | 70 | 0.15 | 467 | 7.5 |
| 2020 [14] | SiPh | 1550 | 8×8 | 8.9×2.2 | 0.92×0.32 | 66 | 8.8 |
| 2019 [70] | SiPh | 1550 | 128 (2D) | 16×16 | 0.8×0.8 | 400 | 12 |
| 2018 [62] | SiPh | 1550 | 1024 | 22.5 | 0.03 | 750 | 9 |
| 2016 [71] | SiPh | 1260 – 1360 | 128 | 80 | 0.14 | 570 | - |
| 2014 [9] | SiPh | 1480 – 1580 | 16 | 19.6 | 1.1 | 17 | 10 |
| 2014 [72] | SiPh | 1550 | 16 | 51 | 3.3 | 15 | - |
| 2011 [63] | SiPh | 1555 | 16 | 14 | 1.6 | 8 | 10 |

[a] Results reported for 1064 nm. [b] Extracted from reported figure.

## 5. Conclusion

This work has successfully demonstrated a 16-channel GaAs-based OPA with 0.92° beamwidth, 15.3° grating-lobe-free steering range, and 12 dB SLL in close agreement with theoretical results. Individual phase modulators tested from 980 nm to 1360 nm demonstrate single-sided $V_\pi \cdot L$ modulation efficiency of 1.23 V·cm and better, under 0.5 dB RAM, less than 20 µW DC electrical power, and greater than 770 MHz electro-optical bandwidth when mounted on a carrier and PCB.

Efforts are underway to integrate gain with this epitaxy. Early designs include an on-chip 1030 nm distributed Bragg reflector laser source and semiconductor optical amplifiers near the output for power boosting, forming a high emission power fully monolithic integrated OPA with no external optical components. Future research will provide alternate wavelength designs and explore two-dimensional output.

**Funding.** No funding agency.

**Disclosures.** The authors declare no conflicts of interest.

**Acknowledgements.** The authors acknowledge Demis John, Paul Verrinder, and Lei Wang for technical discussion, and funding support from MKS Instruments. A portion of this work was performed in the UCSB Nanofabrication Facility, an open access laboratory.

**Data availability.** Data underlying the results presented in this paper are not publicly available at this time but may be obtained from the authors upon reasonable request.